\documentclass[twocolumn,showpacs,preprintnumbers,amsmath,amssymb,prl]{revtex4}

\usepackage{graphicx}
\usepackage{dcolumn}
\usepackage{bm}
\usepackage{color}

\newcommand{\ttise}{1\textit{T}-TiSe$_2$}

\newcommand\T{\rule{0pt}{2.6ex}}
\newcommand\B{\rule[-1.2ex]{0pt}{0pt}}

\begin{document}

\preprint{APS/123-QED}
\bibliographystyle{prsty}

\title{Exciton condensation driving the periodic lattice distortion of \ttise}

\author{C. Monney$^{1,2}$}
\author{C. Battaglia$^3$}
\author{H. Cercellier$^4$}
\author{P. Aebi$^1$}
\email{philipp.aebi@unifr.ch}
\author{H. Beck$^1$}

\affiliation{%
$^1$ D\'epartement de Physique and Fribourg Center for Nanomaterials, Universit\'e de Fribourg, CH-1700 Fribourg, Switzerland\\
$^2$ Research Department Synchrotron Radiation and Nanotechnology, Paul Scherrer Institut, CH-5232 Villigen PSI, Switzerland\\
$^3$ Ecole Polytechnique F\'ed\'erale de Lausanne, Institute of Microengineering, Photovoltaics and Thin Film Electronics Laboratory, CH-2000 Neuch\^atel, Switzerland\\
$^4$ Institut N\'eel, CNRS-UJF, BP 166, 38042 Grenoble, France
}%

\date{\today}

\begin{abstract}
We address the lattice instability of \ttise\ in the framework of the exciton condensate phase. We show that, at low temperature, condensed excitons influence the lattice through electron-phonon interaction. It is found that at zero temperature, in the exciton condensate phase of \ttise, this exciton condensate exerts a force on the lattice generating ionic displacements comparable in amplitude to what is measured in experiment. This is thus the first quantitative estimation of the amplitude of the periodic lattice distortion observed in \ttise\, as a consequence of the exciton condensate phase.
\end{abstract}


\maketitle

In a semimetallic or semiconducting system exhibiting a small electronic band overlap or gap, the Coulomb interaction, when poorly screened, leads to the formation of bound states of holes and electrons, called excitons. If their binding energy $E_B$ is larger than the gap, they may \textit{spontaneously} (without any optical pumping) condense at low temperature and drive the system into a new ground state with exotic properties. This new ground state, called the excitonic insulator phase, has been theoretically predicted in the 1960s \cite{JeromeBasis}. 
Since that time, among the numerous systems proposed for the realization of this peculiar phase, only a few turn out to be serious candidates \cite{WakiEI,BucherEI,RiceEI,BismuthEI,YH3EI}. 
Recently we investigated the charge density wave (CDW) system \ttise\ with angle-resolved photoemission spectroscopy (ARPES), favouring the excitonic insulator phase scenario as the origin of the CDW phase \cite{CercellierPRL,MonneyPRB}, as suggested earlier \cite{WilsonComm,WilsonReview}. 
Furthermore, a superconducting phase has been discovered in this material upon copper intercalation \cite{Morosan} and pressure \cite{ForroSupra}. This produces a very interesting phase diagram reminiscent of the one of the iron pnictides, in the sense that a density wave phase gives way to a superconducting dome upon chemical intercalation. The nature of the competition between the ordered (CDW) phase and the superconducting dome is of central interest. Indeed Sawatzky {\it et al.} argued that the large electronic polarizabilites of pnictides and chalcogenides, which can be interpreted as the consequence of virtual excitonic excitations, may be at the origin of the superconducting phase of this new class of materials, reminding the model of Little developped for organics \cite{SawatzkySupra,LittleOrgan}.

The quasi two-dimensional material \ttise\ undergoes a phase transition towards a commensurate 2x2x2 CDW phase below the critical temperature $T_c\simeq 200$K. A weak periodic lattice distortion (PLD) accompanying the CDW (which requires only electronic degrees of freedom) has been measured, involving small ionic displacements $< 0.1$ \AA\ \cite{DiSalvoSuperLatt}. The occurence of this PLD lead Hughes to suggest a band Jahn-Teller effect as the driving force of the CDW in \ttise\ \cite{HughesBJT}. In this respect, Motizuki and coworkers, based on a tight-binding (TB) fit to the band structure calculated by Zunger and Freeman \cite{nesting} found that, by optimizing electronic vs elastic energy, the observed CDW is realized for an ionic displacement very close to the measured one \cite{SYMtotalenergy}. However, the small ionic displacements in comparison with the high spectral weight carried by the backfolded bands as observed by ARPES supports rather an electronic origin of the CDW \cite{CercellierPRL}.

\begin{figure}
\centering
\includegraphics[width=5cm]{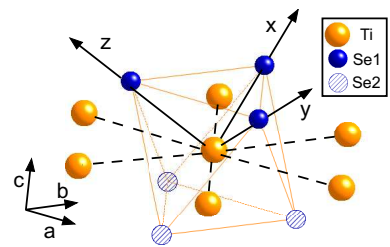}
\caption{\label{fig_1} Atomic structure of \ttise, where the Se (Se1 and Se2) atoms are in (not exactly regular) octahedral coordination around the Ti atoms. The orthonormal axis system points from the center Ti atom towards the neighbouring Se atoms (with a slight deviation of 1.6$^\circ$). The crystallographic axes $a,b,c$ are also shown.
}
\end{figure}

In this perspective, it is crucial to know whether ionic displacements of a reasonable amplitude may appear at all as a consequence of exciton condensation in the low temperature phase. Here, we address this question and study the influence of an exciton condensate on the lattice. 

First, we derive the electron-phonon coupling in the framework of the TB formalism. Focusing particularly on the valence and conduction electrons, we derive a formula relating the ionic displacements to the presence of an exciton condensate, the amplitude of which is directly related to the order parameter characterizing the low temperature phase. Applying this formula to the case of \ttise, we calculate the amplitude of ionic displacements. We find values similar to those obtained from experiment. This demonstrates that the exciton condensate phase, as a possible origin of the CDW phase of \ttise, can also account for the PLD.

The TB formalism for the electronic band structure and the coupling to the lattice described below is similar to that developed by Yoshida and Motizuki \cite{YMsuscept,SYMtotalenergy}. It is applied to the structure of \ttise, which consists of planes of Ti atoms forming a triangular lattice. Each of these Ti atoms is in octahedral coordination with its six neighbouring Se atoms. Then the crystal consists of a regular stacking of such Se-Ti-Se layers along the $c$ direction. For the TB calculations of the present study, a cluster of atoms centered around one Ti atom will be considered. The orthonormal axis system is as shown in Fig. \ref{fig_1}. In our calculations, for the Ti atoms, we include the five $3d$-orbitals of $xy$, $yz$, $xz$, $x^2-y^2$ and $3z^2-r^2$ symmetry and the three $4p$-orbitals for each of the two Se atoms (Se1 and Se2), of $x$, $y$ and $z$ symmetry. In total we have 11 orbitals. The TB electronic Hamiltonian then reads
\begin{eqnarray}\label{eqn_H0}
H_{el}=\sum_{ll'}\sum_{\mu\nu}\sum_{\alpha\beta}J_{\alpha\beta}(\vec{R}_l-\vec{R}_{l'}+\vec{\tau}_\mu-\vec{\tau}_\nu)\nonumber\\
\times\sum_{\vec{k}\vec{k}'n n'}{\rm e}^{-i\vec{k}\cdot\vec{R}_l}T^*_{\alpha\mu,n}(\vec{k}){\rm e}^{i\vec{k}'\cdot\vec{R}_{l'}}T_{\beta\nu,n'}(\vec{k})c^\dagger_{n}(\vec{k})c_{n'}(\vec{k}').
\end{eqnarray}
Here $\vec{R}_l,\vec{R}_{l'}$ are vectors of the Bravais lattice and $\vec{\tau}_\mu,\vec{\tau}_\nu$ are the positions of the ions (Ti, Se1 or Se2) labelled $\mu,\nu$ inside the unit cell. The indices $\alpha,\beta$ label the 11 orbitals and $n$ is the index of the bands in which the operators $c_n^\dagger$ create electrons. The transfer matrix $J$, with the eigenvectors $T$, consists here only of two-center integrals for simplicity.

We now introduce ionic displacements of the form
\begin{eqnarray}\label{eqn_atomdispl}
\vec{u}_{l\mu}=\frac{1}{\sqrt{M_\mu}}\sum_{\vec{q},\lambda}{\rm e}^{i\vec{q}\cdot\vec{R}_l}\vec{e}(\mu,\vec{q},\lambda)Q(\vec{q},\lambda)=\sum_{\vec{q},\lambda}{\rm e}^{i\vec{q}\cdot\vec{R}_l}\vec{u}_\mu(\vec{q},\lambda)
\end{eqnarray}
where $M_\mu$ is the mass of the ion labelled $\mu$, $\vec{e}$ a polarization vector and $Q$ the normal coordinate of the phonons. Here $\vec{u}_\mu(\vec{q},\lambda)$ is the ionic displacement for the atom labelled $\mu$ (Ti, Se1 or Se2) and associated to a particular mode $\vec{q},\lambda$. Equation \ref{eqn_atomdispl} provides us with a direct way to compute the amplitude of the displacement of each ion, once we get a value for $Q$. This is our goal in the next paragraphs. After introducing the small ionic displacements $\vec{u}_{l\mu}$ in the argument of the transfer matrix $J$ in equation \ref{eqn_H0}, we expand $J$ to first order in $\vec{u}_{l\mu}$ to deduce the electron-phonon interaction
\begin{eqnarray}\label{eqn_elattHam}
H_{el-ph}=\sum_{nn'}\sum_{\vec{k}\vec{q},\lambda} g_{nn'}(\vec{k},\vec{q},\lambda)c^\dagger_{n}(\vec{k})c_{n'}(\vec{k}-\vec{q})Q(\vec{q},\lambda)
\end{eqnarray}
where the electron-phonon coupling constant
\begin{eqnarray}\label{eqn_elattcpl}
g_{nn'}(\vec{k},\vec{q},\lambda)=\sum_{\vec{\rho}}\sum_{\mu\nu}\sum_{\alpha\beta}T^*_{\alpha\mu,n}(\vec{k})\frac{dJ_{\alpha\beta}}{d\vec{x}}\bigg\vert_{\vec{x}=\vec{\rho}+\vec{\tau}_\mu-\vec{\tau}_\nu}   \nonumber\\ \times  T_{\beta\nu,n'}(\vec{k}-\vec{q})
 {\rm e}^{-i\vec{k}\cdot\vec{\rho}} \cdot \left[ \vec{e}(\mu,\vec{q},\lambda){\rm e}^{i\vec{q}\cdot\vec{\rho}} -\vec{e}(\nu,\vec{q},\lambda) \right],
\end{eqnarray}
with $\vec{\rho}=\vec{R}_l-\vec{R}_{l'}$, involves the derivatives of the transfer matrix $dJ_{\alpha\beta}/d\vec{x}$.

In what follows, we focus on the influence of excitons. In our simplified model of the band structure of \ttise, we essentially consider the topmost valence band having its maximum at the center of the Brillouin zone ($\Gamma$ point) and the three symmetry equivalent conduction bands at the border of the Brillouin zone ($L$ points), whose extrema are separated from $\Gamma$ by the wave vectors $\vec{w}_i$ ($i=1,2,3$) \cite{MonneyPRB}. These excitons are composed of holes created by $a(\vec{k})$ in the valence band (near its maximum) with a wave vector $\vec{k}$ and electrons created by $b_i^\dagger(\vec{k})$ in the conduction band $i$ (near its minimum) with wave vector $\vec{k}+\vec{w}_i$. Thus the sum over the band indices $n,n'$ in equation \ref{eqn_elattHam} is restricted to terms mixing $a$ and $b$ operators only, so that
\begin{eqnarray}\label{eqn_exclattHam}
H_{el-ph}&=&\sum_{i}\sum_{\vec{k}\vec{q},\lambda}Q(\vec{q},\lambda)g_{ab_i}(\vec{k},\vec{q},\lambda)a^\dagger(\vec{k})b_{i}(\vec{k}-\vec{w}_i-\vec{q})\nonumber\\ && + Q(\vec{q},\lambda)g_{b_ia}(\vec{k},\vec{q},\lambda)b^\dagger_{i}(\vec{k}-\vec{w}_i)a(\vec{k}-\vec{q})\nonumber\\
&=&\sum_{i}\sum_{\vec{p},\lambda}Q(-\vec{w}_i,\lambda)g_{ab_i}(\vec{p},-\vec{w}_i,\lambda)a^\dagger(\vec{p})b_{i}(\vec{p})\nonumber\\ && + Q(\vec{w}_i,\lambda)g_{b_ia}(\vec{p}+\vec{w}_i,\vec{w}_i,\lambda)b^\dagger_{i}(\vec{p})a(\vec{p}).\nonumber
\end{eqnarray}
We considered only $\vec{q}=-\vec{w}_i$ in the term involving $g_{ab_i}$ and $\vec{q}=\vec{w}_i$ in that involving $g_{b_ia}$ (together with the substitution $\vec{p}=\vec{k}-\vec{w}_i$), restricting ourselves to the scattering between the extrema of the bands. 
Then, averaging $\langle H_{el-ph} \rangle_{el}$ to lowest order over the electronic degrees of freedom yields the contribution of condensed excitons to the phonon Hamilonian
\begin{eqnarray}
\langle H_{el-ph} \rangle_{el} &=&\sum_{i}\sum_{\vec{p},\lambda}Q(-\vec{w}_i,\lambda)g_{ab_i}(\vec{p},-\vec{w}_i,\lambda)\langle a^\dagger(\vec{p})b_{i}(\vec{p})\rangle\nonumber\\ && + Q(\vec{w}_i,\lambda)g_{b_ia}(\vec{p}+\vec{w}_i,\vec{w}_i,\lambda)\langle b^\dagger_{i}(\vec{p})a(\vec{p})\rangle.\nonumber
\end{eqnarray}
In analogy to the BCS-theory, the averages $\langle b_i^\dagger a\rangle$ are related to anomalous Green's functions $F_i(\vec{p},\tau)=(-i)\langle T b_i^\dagger(\vec{p},\tau)a(\vec{p})\rangle$ (introduced in reference \cite{MonneyPRB}), so that the previous equation becomes
\begin{eqnarray}
\langle H_{el-ph} \rangle_{el} &=&i\sum_{i}\sum_{\vec{p},\lambda}Q(-\vec{w}_i,\lambda)g_{ab_i}(\vec{p},-\vec{w}_i,\lambda)F_i^\dagger(\vec{p},0)\rangle\nonumber\\ && + Q(\vec{w}_i,\lambda)g_{b_ia}(\vec{p}+\vec{w}_i,\vec{w}_i,\lambda)F_i(\vec{p},0)\nonumber\\
&=&H_{ph-x}.\nonumber
\end{eqnarray}
This exciton-phonon Hamiltonian $H_{ph-x}$ can be further simplified using the inversion symmetry of the system to replace $-\vec{w}_i$ by $\vec{w}_i$ and using also the property $F_i(\vec{p},0)=F_i^\dagger(\vec{p},0)$, giving
\begin{eqnarray}\label{eqn_Ffct}
H_{ph-x} &=&i\sum_{i,\lambda}Q(\vec{w}_i,\lambda)\sum_{\vec{p}}F_i(\vec{p},0)\nonumber\\ && \times\left[ g_{ab_i}(\vec{p},\vec{w}_i,\lambda) + g_{b_ia}(\vec{p}+\vec{w}_i,\vec{w}_i,\lambda)\right]. \nonumber 
\end{eqnarray}
From this last equation, the equilibrium condition for the lattice in the presence of a condensate of excitons, $\partial(H_{ph,0}+H_{ph-x})/\partial Q(\vec{w}_i,\lambda)=0$, leads to an expression for the normal coordinate of the phonons $Q$ caused by the exciton condensate
\begin{eqnarray}\label{eqn_normalcoord}
Q(\vec{w}_i,\lambda)=\frac{i\sum_{\vec{p}}F_i(\vec{p},0) \left[ g_{ab_i} + g_{b_ia} \right]}{\omega^2(\vec{w}_i,\lambda)}
\end{eqnarray}
where $H_{ph,0}=(1/2)\sum_{i,\lambda}\omega^2(\vec{w}_i,\lambda)Q^*(\vec{w}_i,\lambda)Q(\vec{w}_i,\lambda)$  is the bare Hamiltonian of the lattice (in the absence of the exciton condensate). By using equation \ref{eqn_atomdispl} we can relate $Q$ to the resulting ionic displacements.
\\

Having now an analytical formula for the ionic displacements thanks to equations \ref{eqn_atomdispl}, \ref{eqn_elattcpl}, \ref{eqn_normalcoord}, we can look for the necessary numerical parameters for the final computation. We start with the transfer matrix $J_{\alpha\beta}$. According to Slater and Koster \cite{SlaterTB}, its elements are computed as a combination of direction cosines (defining the direction of the bonds joining two atoms) and transfer integrals (specific to the orbitals involved). In our case, these transfer integrals are determined by fitting a band structure computed with density functional theory (DFT). This first-principles band structure has been calculated using the full potential augmented plane wave plus local orbitals (APW+lo) method with the generalized gradient approximation in the parametrization of Perdew, Burke and Ernzerhof \cite{PBE}, in the local density approximation, as implemented in the WIEN2K software package \cite{Wien2k}. The numerical and unit cell parameters \cite{paramsWien,paramsWienLatt} are similar to those used in reference \cite{Jishi}.

\begin{figure}
\centering
\includegraphics[width=7cm]{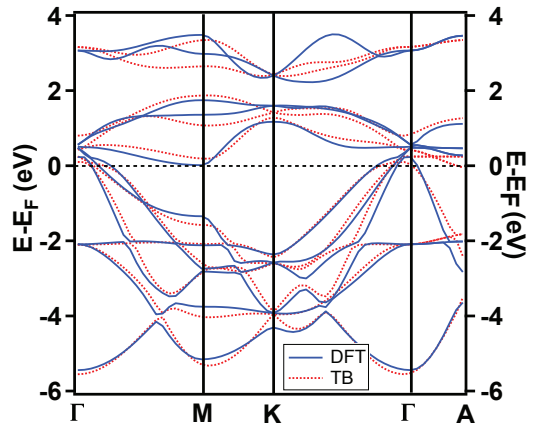}
\caption{\label{fig_2}
Comparison of the band structure of \ttise\ calculated with density functional theory and its fit within a TB approach.}
\end{figure}

The resulting DFT band structure is plotted in Fig. \ref{fig_2}. The goal of this DFT calculation (such a DFT band structure is discussed in detail in references \cite{Jishi,FangLDA}) is to offer a basis for the TB fit we perform for determining the transfer integrals. A least-square fit method was used on our DFT band structure along the high symmetry directions $\Gamma MK\Gamma A$ and the obtained transfer integrals are summarized in Tab. \ref{tab_transfint} \cite{overlapintegrals}. The TB fit of the DFT band structure is good for states below the Fermi energy $E_F$ but becomes poor for states above $E_F$.
This TB parametrization allows to build an effective Hamiltonian which can be diagonalized (at each $\vec{k}$ point) to provide the eigenvectors $T_{\alpha\mu,n}(\vec{k})$. The derivatives of the transfer matrix $dJ_{\alpha\beta}/d\vec{x}$, appearing in the electron-phonon coupling $g_{nn'}(\vec{k},\vec{q},\lambda)$, lead to derivatives of the direction cosines and derivatives of the transfer integrals. The latter are evaluated as in Ref. \cite{SYMtotalenergy}. 

According to the experimental result of Di Salvo {\it et al.} \cite{DiSalvoSuperLatt}, we fix the phonon polarization vectors $\vec{e}(\mu,\vec{q},\lambda)$ involved in the CDW to the direction perpendicular to their respective $\vec{q}$ vector, lying in the $ab$ basal plane. This way, only the transverse phonon mode $\lambda_{tr}$ will be considered in the following calculations. 
\begin{table}[ht]
\caption{Transfer integrals (in eV) for \ttise, obtained from TB fits to first-principles band structure ($\varepsilon_p,\varepsilon_{d\varepsilon},\varepsilon_{d\gamma}$ are the orbital energies) \cite{difftransint}.}
\begin{tabular}{cccc}
\hline\hline
\T\B Transfer integral & Energy  & Transfer integral & Energy   \\ 
\hline
\T
$t(pp\sigma)$ & 0.77  & $t(pd\pi)$ & 0.70  \\
$t(pp\pi)$ & -0.054  & $t(pp\sigma)_2$ & 0.63  \\
$t(dd\sigma)$ & -0.35  & $t(pp\pi)_2$ & -0.028  \\
$t(dd\pi)$ & 0.074  & $t(pp\sigma)_3$ & 0.61  \\
$t(dd\delta)$ & -0.049  & $t(pp\pi)_3$ & -0.096  \\
$t(pd\sigma)$ & 1.3 & & \B \\
\hline
\T\B $\varepsilon_p=-2.0$ eV & $\varepsilon_{d\varepsilon}=0.74$ eV & $\varepsilon_{d\gamma}=1.2$ eV &  \\
\hline\hline 
\end{tabular}
\label{tab_transfint}
\end{table}

The anomalous Green's function $F_i(\vec{p},\tau=0)$ appearing in equation \ref{eqn_Ffct} is calculated as the Fourier transform of $F_i(\vec{p},z)$ given by (see reference \cite{MonneyPRB})
\begin{eqnarray}
F_i(\vec{p},z)=-\frac{\Delta(z-\varepsilon_c^{i+1}(\vec{p}+\vec{w}_2))(z-\varepsilon_c^{i+2}(\vec{p}+\vec{w}_3))}{\mathcal{D}(\vec{p},z)}\nonumber
\end{eqnarray}
with the denominator being
\begin{eqnarray}
\mathcal{D}(\vec{p},z)&=&(z-\epsilon_v(\vec{p}))\prod_i(z-\epsilon_c^i(\vec{p}+\vec{w}_i))\nonumber\\&-&\sum_i|\Delta|^2\prod_{j\neq i}(z-\epsilon_c^j(\vec{p}+\vec{w}_j)).\nonumber
\end{eqnarray}
Here, the order parameter $\Delta$ describes the intensity of the exciton condensate in the low temperature phase. The functions $\varepsilon_v$ and $\varepsilon_c^i$ describe the dispersions of the valence band and of the three symmetry equivalent conduction bands ($i=1,2,3$) close to their maximum situated at $\Gamma$ and at $L$, respectively. The anomalous Green's function $F_i$ is sensitive to the energies appearing in these dispersions near their extrema. We therefore cannot use the TB dispersions, which are too rough with this respect (however they are essential for the more global treatment needed to obtain the transfer integrals), but we need the formulas for $\varepsilon_v$ and $\varepsilon_c^i$ determined in a previous study from fits to ARPES data \cite{CercellierPRL}. 

Finally, combining equations \ref{eqn_atomdispl} and \ref{eqn_normalcoord}, the amplitude of the ionic displacement for a single $\vec{w}_i$ (and for transverse phonons) gets the following form
\begin{eqnarray}\label{eqn_amplatomdispl}
u_\mu(\vec{w}_i,\lambda_{tr})=\frac{1}{\sqrt{M_\mu}}\frac{\left|\sum_{\vec{p}}F_i(\vec{p},0) \left[ g_{ab_i} + g_{b_ia}\right] \right|}{\omega^2(\vec{w}_i,\lambda_{tr})}.
\end{eqnarray}
We now focus to the particular case of the Ti atoms, so that $M_{Ti}$ describes the Ti atom mass. For the order parameter appearing in the anomalous Green's function, we consider a mean-field like temperature dependence of the form $\Delta(T)=\Delta_0\sqrt{1-(T/T_c)^2}$ with $T_c=200$K, where the zero value $\Delta_0=115$ meV has been determined in our recent temperature dependent ARPES study \cite{MonneyTScan}.
In formula \ref{eqn_amplatomdispl}, we use a value of $\omega(\vec{w}_i,\lambda_{tr})= 6.3$ THz. In fact, it is the value estimated by Holt \textit{et al.} at $T\simeq150$K for the transverse phonon mode, which softens at $L$ at the transition \cite{HoltXRD}. This corresponds to a situation where the ionic displacement $\vec{u}_{l\mu}$ is not too large, such that our first order development of the electron-phonon coupling remains valid. Furthermore, the region close to $T_c$ is avoided, where anharmonicities cannot be neglected in the bare Hamiltonian for the lattice $H_{ph,0}$.

\begin{figure}
\centering
\includegraphics[width=6cm]{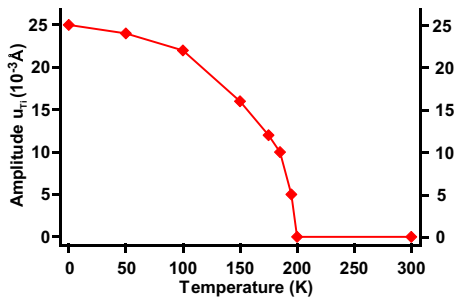}
\caption{\label{fig_3} .
Amplitude of the Ti ionic displacement for a single $\vec{w}_i$, $u_{\text{Ti}}(\vec{w}_i,\lambda_{tr})$ (the diamonds show the calculated values).
}
\end{figure}

Now all the necessary quantities to compute the amplitude of the ionic displacements in equation \ref{eqn_amplatomdispl} are known. Fig. \ref{fig_3} summarizes our numerical results. It shows a clear temperature dependence, following closely the behaviour of the order parameter. Extrapolated to the lowest temperature, it reaches the value of $u_{\text{Ti}}^{\text{theo}}(\vec{w}_i,\lambda_{tr})=0.025$ \AA. Di Salvo \textit{et al.} inferred from neutron diffraction experiments a displacement (also for a single-$\vec{w}_i$) of about $u_{\text{Ti}}^{\text{exp}}(\vec{w}_i,\lambda_{tr})=0.04$ \AA\ at 77K \cite{DiSalvoSuperLatt}. Therefore our value, although being about 60\% of the experimental one, reproduces the measured ionic displacement for Ti atoms within the correct order of magnitude, which is a substantial result, considering the approximations made in this calculation. The uncertainty on $\omega(\vec{w}_i,\lambda_{tr})$ used in equation \ref{eqn_amplatomdispl} may enhance or reduce this value by a factor of 2-3, but it still remains within the correct order of magnitude, in agreement with the main message of this letter.

In conclusion, we addressed the question of the appearance of a periodic lattice distortion in \ttise. Previously we gave strong support for the exciton condensation as a purely electronic mechanism responsible for the CDW phase in this material \cite{CercellierPRL,MonneyPRB}. In this work, we elaborate in a tight-binding formalism a formula for estimating the ionic displacements produced by the presence of this exciton condensate through the electron-phonon coupling. The calculated amplitude of these ionic displacements is, at low temperature, of the same order of magnitude as what is experimentally found. This is thus the first quantitative estimation of the amplitude of the PLD observed in \ttise\, as a \textit{consequence} of the exciton condensate phase. More generally, this result describes quantitatively how an excitonic insulator phase can give rise to a PLD through electron-phonon interaction.

\begin{acknowledgments}
This project was supported by the Fonds National Suisse pour la Recherche Scientifique through Div. II and the Swiss National Center of Competence in Research MaNEP.
\end{acknowledgments}


\begin{thebibliography}{99}
\bibitem{paramsWien}
The APW+lo basis was expanded up to $R_{MT}K_{max}=7$, with $R_{MT}$ the radius of the muffin tin sphere and $K_{max}$ the maximum length of the reciprocal lattice vector. Inside the muffin tin sphere, the $l$ expansion of the nonspherical potential and charge density is carried out up to $l_{max}=10$. The number of $k$ points in the irreducible part of the Brillouin zone is 360. The maximum length of wave vector used for the Fourier expansion of the charge density is $12a_0^{-1}$ ($a_0$ is the Bohr radius).
\bibitem{paramsWienLatt}
The space group for \ttise\ is $P\bar{3}m1$ (164) and the unit cell parameters are $a=b=3.534$ \AA, $c=6.008$ \AA\ (fixed to the experimental values \cite{Morosan}). The reduced coordinates for Ti and for Se atoms are respectively $(0,0,0)$ and $(1/3,2/3,z_{red})$, with $z_{red}=0.265$ determined by minimizing the forces on each atom for fixed $a,b,c$.
\bibitem{overlapintegrals}
In our formulation, no overlap matrix was used to ensure orthogonality of the orbitals. However, this is implicitly included in the transfer matrix \cite{Harrison}.
\bibitem{difftransint}
The values of the transfer integrals in this work are different from those of reference \cite{SYMtotalenergy}, because the origin of energy is different and because no overlap integrals are explicitly considered here (see reference \cite{overlapintegrals}).
\end{thebibliography}
\end{document}